\documentclass[a4paper,12pt]{article}
\usepackage[T2A]{fontenc}
\usepackage[cp1251]{inputenc}
\usepackage[russian,english]{babel}
\usepackage{babelbib}
\usepackage{amsmath,amsfonts,amssymb}
\usepackage{mathtext}
\usepackage[bookmarks=true,unicode=true]{hyperref}
\hypersetup{%
	pdftitle={The Chandler wobble and Solar day},%
	pdfauthor={Kiryan~D.\,G., Kiryan~G.\,V.},%
	pdfsubject={analyzing time series with the step multiple of solar day},%
	pdfkeywords={solar day, time series, gravitation, Chandler wobble, Moon, GPS, UT1},%
	colorlinks={true},urlcolor={blue}}
\usepackage{marvosym}
\usepackage[stable]{footmisc}
\usepackage{url}
\usepackage{breakurl}
\usepackage{setspace}  
\usepackage{misccorr}
\usepackage[vmargin={25mm,30mm},hmargin={37mm,20mm}]{geometry}

\usepackage[draft]{prelim2e}
\usepackage{graphicx}
\usepackage{caption2}
\graphicspath{{images/}}
\usepackage{var7}  

\begin{document}

\vspace*{-15mm}

\begin{center}
    \begin{minipage}{115mm}
    \centerline{\textbf{\Large The Chandler wobble and Solar day}}
    \end{minipage}
\end{center}
\medskip

\centerline{Kiryan~D.\,G. and Kiryan~G.\,V.}
\bigskip

\begin{center}
	\begin{minipage}{110mm}
	\emph{\small Institute of Problems of Mechanical Engineering of RAS,\\
	61 Bolshoy Prospect V.O., 199178, Saint-Petersburg, Russia\\
	e-mail: \textit{diki.ipme@gmail.com}}
	\end{minipage}
\end{center}

%

\section*{}

\begin{center}
\begin{minipage}[h]{140mm}
\begin{spacing}{0.85}
This work supplements the main results given in our paper “\href{http://arxiv.org/abs/1109.4969v4}{\emph{The Chandler wobble is a phantom}}”~\cite{2011arXiv1109.4969K} and refines the reasons for which researchers previously failed in interpreting the physical meaning of observed zenith distance variations. The main reason for the Chandler wobble challenge emergence was that, in analyzing time series with the step multiple of solar day, researchers ignored the nature of the solar day itself. In~addition, astrometric instruments used to measure the zenith distance relative the local normal are, by definition, gravity independent, since the local normal is tangential to the gravitation field line at the observation point. Therefore, the detected zenith distances involve all the instantaneous gravitational field distortions. The direct dependence of the zenith distance observations on the gravitational effect of the Moon’s perigee mass brings us to the conclusion that the Chandler wobble is fully independent of the possible motion of the Earth’s rotation axis within the Earth.
\end{spacing}
\end{minipage}
\end{center}

%

\section{On observation of the zenith distances}

It is commonly recognized that instability of star zenith distances was for the first time revealed by J.~Bradley (1726--1727) and Molyneux (1727--1747) \cite{1972:book:Kulikov,1982:BOOK:Podobed}. In~1840, H.\,I.~Peters pioneered in purposeful detection of the zenith distance variations (latitude variability) by using the most up-to-date at that time optical instruments at the Pulkovo Observatory\footnote{The Main Astronomical Observatory of the Academy of Sciences of the Russia}. Similar observations were being performed at the same observatory from 1863 to 1875 by M.\,O.~Nuren; he was the first who estimated the latitude variation period as $1.2$~year. The issue of making these investigations international was discussed at the International Geodetic Association Congress in 1883 in Rome. The project implementation and practical observations were begun after the Geodetic Association Congress in Salzburg (1888). In~1892, S.~Chandler who has studied and generalized the observations acquired by that time showed that among the latitude variation periods there is one of $400$ to $440$ days~\cite{1902:AJ-v22-p145:Chandler}. By that time, the fact that this phenomenon is caused by motion of the Earth’s rotation axis within the Earth has already been regarded as evident. Soon enough, the following version of the phenomenon interpretation was suggested to the scientific community: the star zenith distance variation (latitude variation) is caused by \emph{free nutation motion} of the Earth’s rotation axis within the Earth~\cite{1892:Newcomb}. This was the first attempt to explain the physical nature of regular variations in the star zenith distance. During the 20th century, other hypotheses were suggested; their common feature was that they were based on the “evidence” but not on proved facts. What was assumed to be evident was that the observed variations in the star zenith distances were caused by the Earth’s rotation axis inclination to the Earth body.

\section{On necessary and sufficient}

Analysis of studies of the zenith distance variations performed during previous years in the scope of national and international programs showed that all those studies have some common features: the star zenith distance variations were observed and interpreted (and are still being observed and interpreted) based on the “evidence”. However, even evident suggestions must be proved; unfortunately, it has not been done~yet.

To prove the existence of the rotation axis motion within the Earth, it could be necessary and sufficient to observe displacement of the Earth’s North and South poles simultaneously, i.e., to simultaneously observe variations in the zenith distance at two points lying on the same meridian on opposite sides of the equator. Such an experiment could provide a researcher clearly understanding the physical meaning of the problem as a whole and, particularly, of the measurements to be performed, with all the data necessary and sufficient to confirm the existence of the Earth’s rotation axis motion within the Earth, i.e., to prove the “evidence”. The “necessary and sufficient” conditions were not fulfilled in any of observations of residual motion of the Earth’s rotation axis; hence, there are no sufficient data for well-reasoned conclusion on the existence of the Earth’s rotation axis motion within the Earth.

This situation cannot be regarded as accidental since neither of the observation programs stipulated direct studying of the Earth’s rotation axis motion. The existence of the rotation axis oscillating-rotating motion within the Earth was postulated as a new entity, namely, free nutation or residual motion. However, all the above-mentioned shows that the fact that seems to be evident is actually an illusion resulting from a chain of misapprehensions.

\section{On the time scale}

In forming the time series, selection of the data sequence time step is important. If a series of observations is constructed with the step equal to the solar day\footnote{The solar day is the $24$-hour time interval between two consecutive upper culminations of the Sun.  Note that, measured in angles, the solar day exceeds $2\pi$ because of the Earth’s rotation about the Sun.}, this means that the Earth’s (and Observer $\mathbf{A}$) self rotation about the axis by $2\pi$ relative to stars is supplemented by rotation in the same direction about the same axis with angular speed~$\omega_\text{year}$:
$$
    \omega_\text{year}=\frac{2\pi}{T_\text{year}}\;,\quad
    T_\text{year} = 365.25~\text{[day]}
$$

Scalar presentation of the total angular speed of the Earth’s self-rotation about the Sun is as follows:
    \begin{equation}\label{eq:var7:OmegaTerra}
            \omega=\omega_\ast + \omega_\text{year}\;,
    \end{equation}
where $\omega$ is the angular speed of the Earth’s self-rotation about the Sun;
$\omega_\ast$ is the angular speed of the Earth’s self-rotation relative to stars\footnote{
The Earth turns by $2\pi$ about its axis relative to stars during $\approx\!23^\text{h}56^\text{m}04^\text{s}$};
$\omega_\text{year}$ is the angular speed of additional Earth’s rotation relative to stars that forms the solar day, namely, the $24$-hour time interval.

\begin{figure}[htb]
        \centering\includegraphics[scale=1.2]{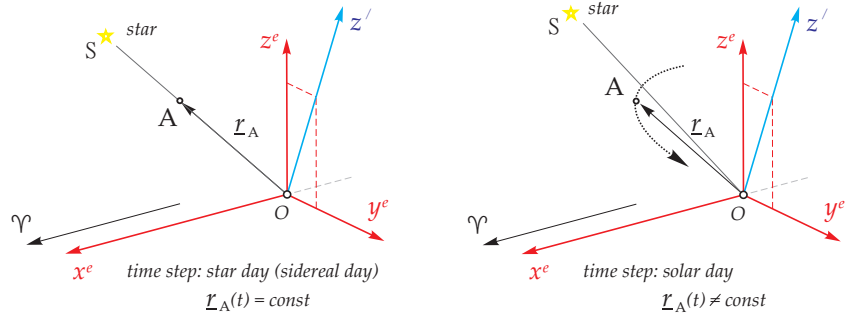}
        \caption{The effect of the time scale selection on the observations interpretation. Axis $\mathbf{O}x^e$ of the Cartesian coordinate system $\mathbf{O}x^ey^ez^e$ is parallel to the vernal equinox line~\Aries\ a in the Earth’s motion (point $\mathbf{O}$) about the Sun. Axis $\mathbf{O}z'$ is the Earth’s self-rotation axis.}
        \label{fig:var7-A-days}
     \end{figure}

Thus, the extra Observer’s rotation that must be taken into account in defining the apparent periods of external perturbations is expressed via angular speeds.  If the time series step is assumed to be equal to the equinoctial day, the Observer will remain motionless with respect to stars. A ray drawn from point $\mathbf{O}$ towards star~$\mathbf{S}$ and passing through observation point~$\mathbf{A}$  will move only plane-parallel but will not rotate. On the contrary, if the measurement interval is equal to or multiple of the solar day, the Observer will move relative to stars with period $T_\text{year}$. This is clearly demonstrated in Fig.~\ref{fig:var7-A-days}.

All the above mentioned brings us to a conclusion that, if we want to get data to be interpreted in processing time series with a step multiple of the solar day, it is necessary to take into account the Earth’s rotation about its axis with the $T_\text{year}$ period.

\section{Certainty of observations}

The plumb-in line is a line tangential to the gravitational field line at the preset point. This means that spatial location of the astronomic instrument zero point changes according to the instantaneous configuration of the Earth’s gravitational field, which makes the instrument gravity-dependent. Hence, the lower are the measured and detected angles and the more accurate and sensitive is the instrument reference system, the higher is the probability that the real Observer of the star zenith distance variations will detect the variations in the spatial location of the astronomic instrument zero point (plumb-in line).

The gravity-dependent astronomic instruments are used to detect the Earth’s self rotation. Observations on the Earth’s daily rotation distorted by instability of the plumb-in line spatial location, as well as data on the postulated Earth’s rotation axis (pole) motion within the Earth, are currently introduced into the Universal Time System (UT1, UT2)~\cite{1971:book:TIME} as relevant corrections, with all the inevitable consequences.

Observation and detection of time moments and small angles with extremely high accuracy in the continuously varying gravitational field of the Earth with gravity-dependent astronomic instruments can lead only to another chain of misapprehensions.

\section{On the \texorpdfstring{\IERS\footnote{%
International Earth Rotation and Reference Systems Service, \url{http://www.iers.org}\\
Earth Orientation Center, \url{http://hpiers.obspm.fr/eop-pc}}}{\IERS} observation series}

\begin{figure}[htb]
        \centering\includegraphics[scale=1.2]{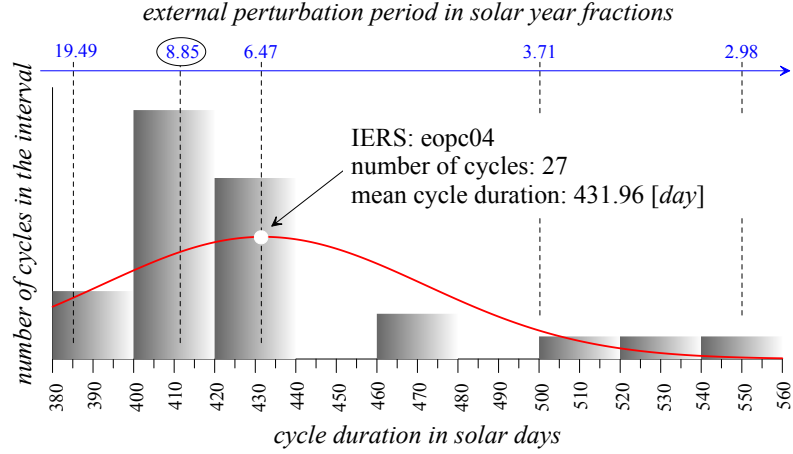}
        \caption{Duration distribution of cycles ($2\pi$) in the “residual motion” of the Earth’s rotation axis
        within the Earth (\IERS data)}
        \label{fig:var7-graphics-histogram2}
     \end{figure}

In processing observations of celestial body zenith distance variations, maximum attention was given to statistical methods, the physical meaning being almost fully ignored. Analysis of observations only by statistical methods, ignoring the phenomenon physics, often pushes the investigation into the metaphysics territory. Fig.~\ref{fig:var7-graphics-histogram2} demonstrates a histogram representing the \IERS data analysis results that are the duration distribution of cycles ($2\pi$) of the postulated residual motion of the Earth’s rotation axis within the Earth.

The maximal frequency of cycles with certain periods in the Earth’s rotation axis “motion” manifests the existence of a process with the period of about $412$~days. In calculating this period, the additional annual rotation of Observer $\mathbf{A}$ was ignored; hence, it would be, evidently, a false step to try to interpret this period and use it in searching for an appropriate physical phenomenon.

The data time series have been constructed from daily mean values of the observed angles, i.e., zenith distances. The daily and multi-daily averaging improved statistical accuracy of calculations. However, the fact that the selected series step, i.e., solar day, is a result of summing two rotations~\eqref{eq:var7:OmegaTerra} was missed. This fact must be taken into account in interpreting the observation results. The Observer self-rotation with the $T_\text{year}$ period changes the periods of external perturbing factors detected in observations. For instance, the Moon’s perigee mass\footnote{The Moon’s gravitational effect on the Earth is substituted with the equivalent gravitational effect of a certain body located in the Moon’s perigee~\cite{2011arXiv1109.4969K}. This body mass is derived from the Moon’s total gravitational effect on the motionless Earth during a lunar cycle ($\approx\!28$~days).} rotates counterclockwise with period~$T_\text{perigee}\!=\!8.85\:\text{[year]}\!=\!3232.46\:\text{[days]}$ about the Earth’s center of gravity and disturbs the Earth’s gravitational field. In the scope of classical mechanics, the period of variation in the plumb-in line spatial location caused by this perturbation, which is detected by the Earth’s Observer, can be defined in year fractions as follows:
    \begin{equation}\label{eq:var7:TS365}
        T=\frac{T_\text{perigee} \cdot T_\text{year}}{T_\text{perigee} - T_\text{year}}=
            \frac{8.85 \cdot 1}{8.85 - 1}\approx 1.13\;\text{year}\;\;\text{or}\;
            \approx 412\;\text{days}
    \end{equation}

The same period of the plumb-in line gravitational perturbation defined on the time scale with the step not multiple of the solar day will be different:
    \begin{equation}\label{eq:var7:TS1}
        T=\frac{T_\text{perigee} \cdot T_\text{day}}{T_\text{perigee} - T_\text{day}}=
            \frac{3232.46 \cdot 1}{3232.46 - 1}\approx 1.00031\;\text{day}\;.
    \end{equation}

Consider Fig.~\ref{fig:var7-graphics-histogram2}. The lower (linear) scale represents cycle durations obtained by analyzing the observation data time series with the interval multiple of the solar day. The periods have been computed not taking into account the Observer’s self rotation with the $T_\text{year}$ period. The upper (nonlinear) scale represents the intervals of external perturbation periods. It is adequate to the lower scale of cycle durations, but only after correction for the Observer’s self-rotation with the period of~$T_\text{year}$. Comparing the periods computed ignoring the annual rotation of the Observer with time intervals of possible external perturbations determined with regard to the $T_\text{year}$ rotation, we can reliably determine the interval for searching the external source of natural perturbation of the Earth’s gravitational field for each detected perturbation.

The period of the external (primary) perturbing factor varies from $8$ to $11$ years; due to it, the $412$-day period appears in the observations. The only possible source of this perturbation is the perigee Moon whose period of rotation about the Earth is~$T_\text{perigee}$. Thus, we can state that the $412$-day period is of the natural origin.

\section{Summary}

This paper presents the results of the system approach to considering the problems of observation performance, methods for processing and interpreting the observation results, and the existing gravity dependence of astronomic instruments. The source of the dead-end situation~\cite{1987:SGeo-v9-p419:Runcorn} occurred in trying to reveal the physical nature of the zenith distance variation (latitude variability) was found out. Some mistakes in processing the observation data\footnote{The Observer’s relative rotation with the $T_\text{year}$ period hidden by the time series step multiple of the solar day was ignored.} became a source of wrong interpretation of the observed regular variations in the zenith distance. This means that at some moment at the turn of the 20th century astronomers lost the physical meaning of star zenith distances detection and, moreover, those observations were used to confirm farfetched hypotheses. The scientific community has postulated the results of wrong interpretation of astronomic observations. Moreover, in some geophysics areas the Chandler’s residual Earth’s rotation axis motion within the Earth became an experimental criterion. For instance, validity of the Earth rotation theory is always verified by comparing parameters of the theoretical “Chandler wobbles” of the Earth’s rotation axis with those determined in observations. The observed “Chandler wobbles” of the Earth’s rotation axis are used as an undoubtful criterion in developing the Earth structure models.

The Chandler wobble itself is one of the parameters of a natural phenomenon that has no concern with the Earth’s rotation axis. The Chandler wobbles detected for the rotating Earth are oscillations caused by the gravitational interaction between the mass of the perigee Moon rotating with the $T_\text{perigee}$ period and a gravitating mass on the Earth rotating with the $T_\text{year}$~period. For instance, the gravitational perturbation with the $412$-day period detected by the Observer rotating together with the Earth manifests itself as the plumb-in line deviations (bubble-tube reading variations), atmospheric pressure variations~\cite{2009:article:BAR}, and sea level variations (polar tide). Circadian variations in star zenith distances of the same magnitude as the Chandler wobbles, which were detected by A.\,S.~Vasilyev~\cite{1952:article:Vasiliev}, are of the same origin as the Chandler wobbles, since the zenith distances were observed by using the same gravity-dependent astronomic instruments. Since the time series were constructed with the step not multiple of the solar day, the zenith distance variation period was not Chandler’s, but circadian.

The question of existence of the Earth’s rotation axis motion within the Earth and of quantitative estimation of that motion characteristics still remains open. Answers might be obtained by studying mass redistribution within the Earth~\cite{1999:IERS-TN25:Geocenter}.


\begin{spacing}{0.9}
\bibliography{../../_bib/var,../../_bib/kiryan}
\bibliographystyle{plain}
\end{spacing}

\end{document}